\documentclass[12pt,aps,prb,jmp]{revtex4}

\usepackage{graphicx}
\usepackage{dcolumn}
\usepackage{bm}
\usepackage{amsmath}
\usepackage{amssymb}

\begin{document}

\title{Perturbation Theory for Plasmonic Modulation and Sensing}

\author{Aaswath Raman}
 \email{aaswath@stanford.edu}  
 \affiliation{Department of Applied Physics, Stanford University, Stanford, CA, 94305 USA}
\author{Shanhui Fan}
\email{shanhui@stanford.edu}  
 \affiliation{Department of Electrical Engineering, Stanford University, Stanford, CA, 94305 USA}

\date{\today}

\begin{abstract}
We develop a general perturbation theory to treat small parameter changes in dispersive plasmonic nanostructures and metamaterials. We specifically apply it to dielectric refractive index, and metallic plasma frequency modulation in metal-dielectric nanostructures. As a numerical demonstration, we verify the theory's accuracy against direct calculations, for a system of plasmonic rods in air where the metal is defined by a two-pole fit of silver's dielectric function. We also discuss new optical behavior related to plasma frequency modulation in such systems. Our approach provides new physical insight for the design of plasmonic devices for biochemical sensing and optical modulation, and future active metamaterial applications. 
\end{abstract}

\maketitle

\section{Introduction}
To design active optical devices such as sensors, switches and modulators, one needs to calculate how a small change in refractive index affects the device response function. For active devices based on dielectric structures\cite{{Chow:04},{Mortensen:2008p677},{White:2008p646},{Robinson:2008p740},{DellOlio:2007p695}} described by a frequency-independent dielectric constant distribution $\varepsilon(\mathbf{r})$, the effect of an index change can be understood in terms of a frequency shift $\omega_1$ of the eigenmodes of the system, which is given by first-order perturbation theory as:\cite{{Mortensen:2008p677},{Joannopoulos08}}
\begin{equation}
\omega_1 = -\frac{\omega}{2} \frac{\int d\mathbf{r} \Delta \varepsilon(\mathbf{r}) | \mathbf{E}(\mathbf{r})|^{2}}{\int d\mathbf{r} \varepsilon(\mathbf{r}) | \mathbf{E}(\mathbf{r})|^{2}}
\end{equation}
The numerator in Eq. (1) only has contributions from the perturbed regions as described by $\Delta \varepsilon(\mathbf{r})$. The shift in the eigenfrequency thus depends on the overlap of the modal electric field energy with the perturbed region. 

In recent years, there has been substantial interest in using plasmonics for active devices. While surface-plasmon sensors are already prominent in biochemical sensing applications,\cite{{Homola:2003p1739},{Homola:2006}} remarkable improvements in device performance have been achieved using plasmonic nanostructures and metamaterials.\cite{{Kabashin:2009},{rosenberg:2010},{Alleyne:2007p1671}} Active plasmonic devices have also been implemented for modulation and switching.\cite{{plasmostor},{cai2009},{Fan:2010},{Pala:2008},{Pacifici:2007}} Given these developments, it is essential to develop a general photonic perturbation theory for plasmonic nanophotonic structures to firmly ground the analysis and characterization of next-generation devices. 

Eq. (1), however, is not applicable for plasmonic systems. For example, $\varepsilon(\mathbf{r})$ can be negative in a metal system, and hence directly applying Eq. (1) could lead to a prediction of infinite sensitivity, which is unphysical. Moreover, in plasmonic systems, recent experiments have successfully varied the plasma frequency of the metal as well,\cite{{Shao:2010},{Diest:2010},{Guler:10}} introducing a new degree of freedom that requires formal theoretical treatment.

In this paper, we build upon a formulation recently developed to model the photonic band structure of dispersive material systems,\cite{Raman:2010} to construct a perturbation theory that can predict modal frequency shifts due to changes in the dielectric constants of dispersive systems involving both metals and dielectrics. We also show that a similar perturbation theory can be developed when the plasma frequency of the dispersive metal is modulated, and highlight new physical behavior related to such modulation. Our theory thus allows us to treat variations in key parameters of both metal and dielectric components of plasmonic nanostructures.

The paper is organized as follows. In Section II we review the photonic band theory previously developed to describe dispersive metamaterials and plasmonic nanostructures. We develop a general perturbation theory in Section III to describe the effect of small parameter changes on the optical modes of dispersive nanostructures. We consider the specific cases of refractive-index variation in Section IV, and plasma frequency variation in Section V, and demonstrate the accuracy of this perturbation theory numerically for both cases by comparing to direct calculations. Finally, we conclude in Section VI.

\section{Review of Dispersive Metamaterial Band Theory}

As the basis for the developments of this paper, we first briefly review the theory recently developed to model the photonic modes of dispersive material systems.\cite{Raman:2010} This theory is applicable for a nanophotonic structure containing a dispersive material with a dielectric function
\begin{equation}
\varepsilon(\omega) = \varepsilon_\infty + \varepsilon_\infty \sum_{n=1}^{N} \frac{\omega_{p,n}^{2}}{\omega_{0,n}^{2} - \omega^{2} + i\omega\Gamma_n}.
\end{equation}

To describe such a dispersive material, for the $n$-th pole in the dielectric function, one introduces a polarization field $\mathbf{P}_n$ and a polarization velocity field $\mathbf{V}_n$, satisfying 
\begin{align}
\frac{\partial \mathbf{P}_n}{\partial t} &= \mathbf{V}_n\\
\frac{\partial \mathbf{V}_n}{\partial t} &= \omega_{p,n}^2 \varepsilon_\infty \mathbf{E} - \omega_{0,n}^2 \mathbf{P}_n - \Gamma_n \mathbf{V}_n
\end{align}
These auxiliary fields\cite{{taflove2000},{Joseph:91}} are then coupled to Maxwell's equations through:
\begin{align}
\frac{\partial \mathbf{H} }{\partial t} & = - \frac{1}{\mu_0} \nabla \times \mathbf{E} \\
\frac{\partial \mathbf{E} }{\partial t} & =  \frac{1}{\varepsilon_\infty}  (\nabla \times \mathbf{H}  - \sum_{n=1}^{N}\mathbf{V}_n)
\end{align}

For steady state, with fields varying as $\exp(i \omega t)$, Eqs. (3)-(6) become
\begin{align}
\label{begfreqeq}
i \omega \mathbf{H} &=  - \frac{1}{\mu_0} \nabla \times \mathbf{E} \\
i \omega \mathbf{E} &=   \frac{1}{\varepsilon_\infty}  (\nabla \times \mathbf{H}  - \sum_{n=1}^{N} \mathbf{V}_n) \\
i \omega \mathbf{P}_n &=  \mathbf{V}_n \\
\label{endfreqeq}
i \omega \mathbf{V}_n &=  \omega_{p,n}^2\varepsilon_\infty \mathbf{E} - \omega_{0,n}^2 \mathbf{P}_n - \Gamma_n \mathbf{V}_n
\end{align}
and thus define an eigenvalue problem for $\omega$.  Eqs. \eqref{begfreqeq}-\eqref{endfreqeq} also define a total energy density
\begin{equation}
\label{eq:total_energy}
W_0 = \frac{1}{4} (\varepsilon_\infty |\mathbf{E}|^2 + \mu_0 | \mathbf{H} |^2 ) + \sum_{n=1}^{N} \frac{1}{4 \varepsilon_\infty \omega_{p,n}^2}  (\omega_{0,n}^2 |\mathbf{P}_{n}|^2 + | \mathbf{V}_{n}|^2)
\end{equation}
The spatial integral of Eq. \eqref{eq:total_energy}, which represents the total energy of the system, is conserved when $\Gamma_n = 0$ for all poles. Further, defining $\mathbf{x} = \left ( \mathbf{H},\mathbf{E},\mathbf{P}_1,\mathbf{V}_1, \cdots,\mathbf{P}_N, \mathbf{V}_N \right )^{\mathsf{T}}$, which represents a multi-component vector field varying over the whole space, we can write Eqs. \eqref{begfreqeq}-\eqref{endfreqeq} as
\begin{equation}
\label{maineqn}
\omega \mathbf{A} \mathbf{x}= \mathbf{B} \mathbf{x},
\end{equation}
where $\mathbf{A} = \mathrm{diag} \left ( \mu_0, \varepsilon_\infty,  \cdots, \omega_{0,N}^2/\omega_{p,N}^2\varepsilon_\infty,1/\omega_{p,N}^2\varepsilon_\infty \right )$ and 
\begin{equation}
\mathbf{B} = \left ( \begin{array}{ccccc}
0 & i \nabla \times & \cdots & 0 & 0\\
-i \nabla \times & 0 & \cdots & 0 & i\\
\vdots & & \ddots & & \vdots \\
0 & 0 & \cdots & 0 & -i\frac{\omega_{0,N}^2}{\omega_{p,N}^2\varepsilon_\infty}\\
0 & -i & \cdots & i \frac{\omega_{0,N}^2}{\omega_{p,N}^2\varepsilon_\infty} & i \frac{\Gamma_N}{\omega_{p,N}^2\varepsilon_\infty} \end{array} \right ).
\end{equation}

Finally, defining $\mathbf{y = \sqrt{A}x}$, we can re-write this as an eigenvalue equation for $\omega$:
\begin{equation}
\label{maineqnherm}
\omega \mathbf{y} = \left ( \sqrt{\mathbf{A}} \right )^{-1} \mathbf{B} \left ( \sqrt{\mathbf{A}} \right )^{-1} \mathbf{y}.
\end{equation}
For the lossless case, $\Gamma_n = 0$, Eq. \eqref{maineqnherm} becomes a Hermitian eigenvalue equation which results in an orthogonality condition:
\begin{equation}
\label{eq:ortho}
\int   d \mathbf{r} \left[  \frac{1}{4} (\varepsilon_{\infty} \mathbf{E_a^*} \cdot \mathbf{E_b} + \mu_0 \mathbf{H_a^*} \cdot \mathbf{H_b} ) + \sum_{n=1}^{N}  \frac{1}{4 \varepsilon_\infty \omega_{p,n}^2} ( \omega_{0,n}^2 \mathbf{P}_{\mathbf{a},n}^\mathbf{*} \cdot \mathbf{P}_{\mathbf{b},n} + \mathbf{V}_{\mathbf{a},n}^\mathbf{*} \cdot \mathbf{V}_{\mathbf{b},n}) \right] = \delta_{ab}
\end{equation}

\section{Perturbation Theory}

In this section, we develop a general perturbation theory based on the generalized eigenvalue equation for the photonic bands of dispersive nanostructures, Eq. \eqref{maineqn}. We start from the unperturbed system
\begin{equation}
\label{unperturb}
\omega_0 \mathbf{A}_0 \mathbf{x}_0 = \mathbf{B}_0 \mathbf{x}_0
\end{equation}
In the presence of a perturbation, the system matrices become $\mathbf{A} = \mathbf{A}_0 + \mathbf{A}_1$, and $\mathbf{B} = \mathbf{B}_0 + \mathbf{B}_1$, and as a result, we have 
\begin{align}
\label{firstorder}
(\omega_0 + \omega_1) (\mathbf{A}_0 + \mathbf{A}_1 )(\mathbf{x}_0 + \mathbf{x}_1 ) =  (\mathbf{B}_0 + \mathbf{B}_1 )(\mathbf{x}_0 + \mathbf{x}_1 ) 
\end{align}
Using Eq. \eqref{unperturb} and keeping only first-order terms in Eq. \eqref{firstorder}, we have
\begin{align}
\label{perturbstep1}
\omega_{0}\mathbf{A}_{0}\mathbf{x}_{1} + \omega_{1}\mathbf{A}_{0}\mathbf{x}_{0} + \omega_{0}\mathbf{A}_{1}\mathbf{x}_{0} = \mathbf{B}_{0}\mathbf{x}_{1} + \mathbf{B}_{1}\mathbf{x}_{0}
\end{align}
Eq. \eqref{unperturb}, in its most general form, describes a lossy system and cannot be written as a Hermitian eigenvalue problem. Thus, to calculate $\omega_1$ we also need to determine the left eigenvector $\mathbf{z}_0$ that satisfies
\begin{equation}
\label{lefteig}
\omega_0 \mathbf{z}_0 \mathbf{A}_0 = \mathbf{z}_0 \mathbf{B}_0
\end{equation}
Multiplying $\mathbf{z}_0$ through Eq. \eqref{perturbstep1} and solving for $\omega_1$, we find:
\begin{align}
\label{generalperturb}
\omega_1 = \frac{\mathbf{z}_{0}\mathbf{B}_{1}\mathbf{x}_{0} - \omega_{0}\mathbf{z}_{0}\mathbf{A}_{1}\mathbf{x}_{0}}{\mathbf{z}_{0}\mathbf{A}_{0}\mathbf{x}_{0}}
\end{align}
Eq. \eqref{generalperturb} is the main result of this paper. In the following sections, we apply this equation to two relevant examples of perturbations in plasmonic systems. 

\section{Application I: Dielectric Refractive-Index Modulation}
In this section we consider the specific case of a small change ($\Delta\varepsilon (\mathbf{r})$) in the dielectric constant of a dielectric region, in a nanostructure consisting of both metal and dielectric regions. The metal region is assumed to be unperturbed. In this case, the perturbation takes the form
\begin{equation}
\mathbf{A}_1 = \mathrm{diag}(0,\Delta \varepsilon(\mathbf{r}),\cdots,0,0),
\end{equation}
while $\mathbf{B}_1 = 0$. We now determine the change in modal frequency for the cases when the metal in the metal-dielectric nanostructure is lossless and lossy.

\subsection{Lossless case, $\Gamma_n = 0$}

For the lossless case Eq. \eqref{maineqn} both $\mathbf{A}_0$ and $\mathbf{B}_0$ are Hermitian. In this case, from Eq. \eqref{lefteig}, we have $\mathbf{z}_0 = \mathbf{x}_0^{\dagger}$. Thus, Eq. \eqref{generalperturb} reduces to
\begin{align}
\label{refindlossless}
\omega_1 = - \omega_{0}\frac{\mathbf{x}_{0}^{\dagger}\mathbf{A}_{1}\mathbf{x}_{0}}{\mathbf{x}_{0}^{\dagger}\mathbf{A}_{0}\mathbf{x}_{0}} =  -\omega_{0} \frac{\int d\mathbf{r} \Delta \varepsilon(\mathbf{r}) | \mathbf{E}(\mathbf{r})|^{2}}{\int d\mathbf{r}~W_{0}}.
\end{align}
For the lossless dispersive system, we thus obtain a result that has the same form as Eq. (1) that is now appropriate for a system with dispersion, provided that we consider the \emph{total} energy density in the system including contributions from the auxiliary mechanical fields. The expression for total energy, Eq. \eqref{eq:total_energy}, has contributions from multiple Lorentz poles and is a multi-pole extension of the energy density expression previously derived by taking electric polarization into account explicitly.\cite{{loudon1970},{ruppin2002}} For the lossless case this reduces to the usual expression for energy density in metals\cite{loudon1970}
\begin{equation}
\label{simplewo}
W_0 = \frac{1}{4}\left [ \frac{d(\omega\varepsilon\varepsilon_{\infty})}{d\omega} \right ] |\mathbf{E}|^{2} + \frac{\mu_0}{4} | \mathbf{H} |^2 .
\end{equation}

\subsection{Lossy case, $\Gamma_n \neq 0$}
For the lossy case, the matrix $\mathbf{B}_0$ in Eq. \eqref{unperturb} is no longer Hermitian and $\mathbf{z}_0 \neq \mathbf{x}_0^{\dagger}$. Thus, Eq. \eqref{generalperturb} in this case reduces to:
\begin{align}
\label{refindlossy}
\omega_{1} = -\omega_{0}\frac{\mathbf{z}_{0}\mathbf{A}_{1}\mathbf{x}_{0}}{\mathbf{z}_{0}\mathbf{A}_{0}\mathbf{x}_{0}}
\end{align}
While no explicit expression analogous to Eq. (1) can be written for the lossy case, Eq. \eqref{refindlossy} still allows one to calculate the frequency shift due to a dielectric refractive index change in the presence of a lossy metal; an important ability in realistic plasmonic sensing schemes.

Eq. \eqref{refindlossy} represents the technically correct way to do perturbation theory, where one needs to determine both the left and right eigenvectors of the general eigenvalue problem. Moreover the denominator in Eq. \eqref{refindlossy} cannot be interpreted as an energy integral. Empirically, on the other hand, we will show numerically that in fact $\mathbf{z}_0 \approx \mathbf{x}_0^{\dagger}$, even for metals with realistic loss parameters, and thus the denominator of Eq. \eqref{refindlossy},
\begin{align}
\label{lossyapprox}
\mathbf{z}_{0}\mathbf{A}_{0}\mathbf{x}_{0} \approx \int d\mathbf{r}~W_{0}
\end{align}
where $W_0$ is the energy density of the mode for the lossy system as defined in Eq. \eqref{eq:total_energy}, which includes contributions from the mechanical auxiliary fields. We note that in a lossy system, when multiple poles are involved, there is no simple relation such as Eq. \eqref{simplewo} that can be used to describe the total energy. Instead, the definition of Eq. \eqref{eq:total_energy}, which explicitly takes into account contributions from the auxiliary mechanical fields, must be used. 

\subsection{Numerical Example}

We now numerically verify the accuracy of the perturbation theory results presented. Motivated by a recent experiment,\cite{Kabashin:2009} we consider as our model system a two-dimensional periodic array of square plasmonic rods in air that are uniform along the third $z$ direction, and numerically solve its eigenmodes. The system has periodicity $a = 130$nm and the rod has a side length of $s = 0.45a = 58.5$nm. 

\begin{figure}[t]
 \begin{center}
   \includegraphics[scale=0.62]{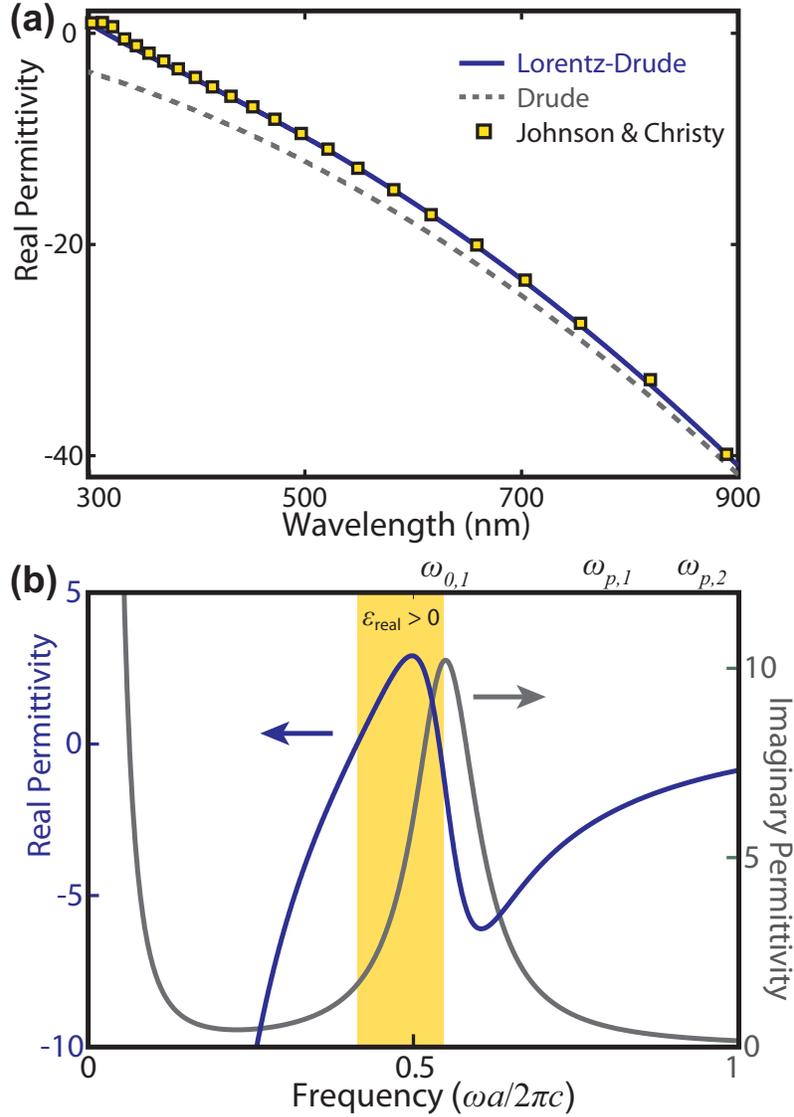}
  \end{center}
  \caption{(a) Comparing tabulated data for the real part of silver's permittivity\cite{jc:1972} against the Drude model and Lorentz-Drude\cite{Drachev:2008p2030} fit used in this paper, at optical wavelengths. (b) The imaginary and real parts of the Lorentz-Drude fit\cite{Drachev:2008p2030} of silver's permittivity over normalized frequencies for $a = 130$nm.}
\end{figure}

\subsubsection{Metal Dielectric Function Fit}

The plasmonic metal's dielectric function is defined by fitting silver's tabulated permittivity\cite{jc:1972} with a Lorentz-Drude model,\cite{Drachev:2008p2030} consisting of a Lorentz pole and a Drude pole [Eq. (2)]. In Fig. 1(a) we see that using the two poles is more accurate than using the Drude model alone for modeling silver's dielectric function at optical frequencies. The Lorentz pole in this fit is defined by $\omega_{0,1} = 0.5526$, $\omega_{p,1} = 0.8196$, and $\Gamma_1 = 0.1195$, and the Drude pole by $\omega_{0,2} = 0$, $\omega_{p,2} = 0.9615$, and $\Gamma_2 = 0.0022$. All frequencies in the numerical examples are normalized to $2\pi c/a$. A plot of the real and imaginary parts of this fit over the relevant normalized frequency range is presented in Fig. 1(b). In the formalism of Section II, the Drude pole has zero resonant frequency, and hence only requires the $\mathbf{V}$-field as its auxiliary field. Thus, in our system, we describe the effects of dispersion in terms of three auxiliary fields: $\mathbf{P}_1$ and $\mathbf{V}_1$ for the Lorentz pole, and $\mathbf{V}_2$ for the Drude pole.  

\begin{figure}[t]
 \begin{center}
   \includegraphics[scale=0.64]{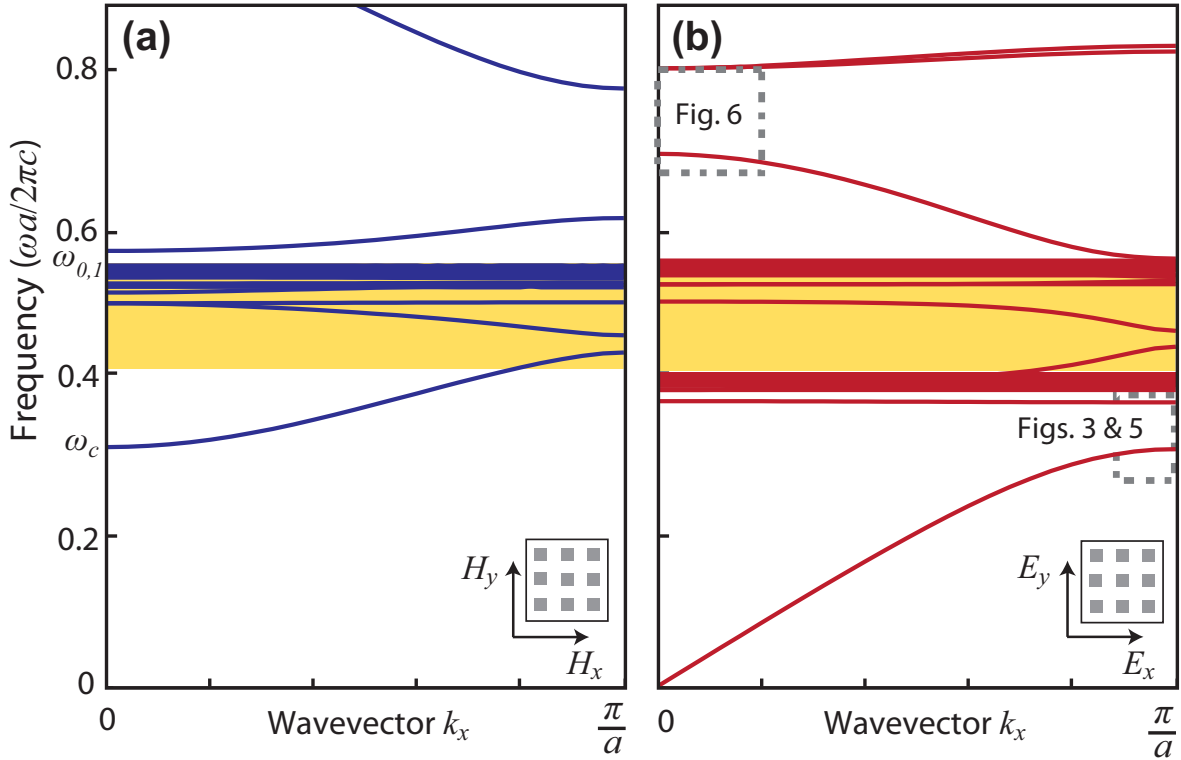}

  \end{center}
  \caption{Computed band structures for a square lattice of 2D plasmonic rods ($s = 0.45a$) in air for the (a) TM and (b) TE polarization between $(k_x = 0, k_y = 0)$ and $(k_x = \pi/a, k_y = 0)$ points. The metal is describe by a Lorentz-Drude fit of silver with the presence of the Lorentz pole at $\omega_{0,1}$ highlighted. The cutoff frequency for TM modes is identified at $\omega_c$, and frequency regions where Re[$\varepsilon$] $> 0$ are highlighted in yellow.}
\end{figure}

\subsubsection{Theory Verification}

We analyze the TM and TE modes (with their electric and magnetic fields respectively polarized along the $z$-dimension) of this system in Fig. 2. These band structures are calculated by implementing a finite-difference spatial discretization of the fields with a Yee grid, and then solving the eigenvalue equation using the Arnoldi method.\cite{Raman:2010} In our finite difference implementation, we have truncated the finite-difference grid appropriately at the metal-air interface to ensure that boundary conditions for the tangential field components are satisfied at these interfaces.\cite{{Fan:1996},{Ferrario:2010}} For the TM case, we find that a stop-band exists below a cutoff frequency $\omega_c = 0.3067$. In the TE case we note the presence of dispersion-less flat bands below $\omega \approx 0.4$ that correspond to surface plasmon modes, and modes that have substantial group velocity which correspond to non-surface modes. For both cases, we also observe the presence of a dense cluster of low group-velocity modes right below $\omega_{0,1}$\cite{Huang03} in the region where Re$[\varepsilon] > 0$, highlighted yellow in Fig. 1(b) and Fig. 2.

\begin{figure}[t]
 \begin{center}
   \includegraphics[scale=0.62]{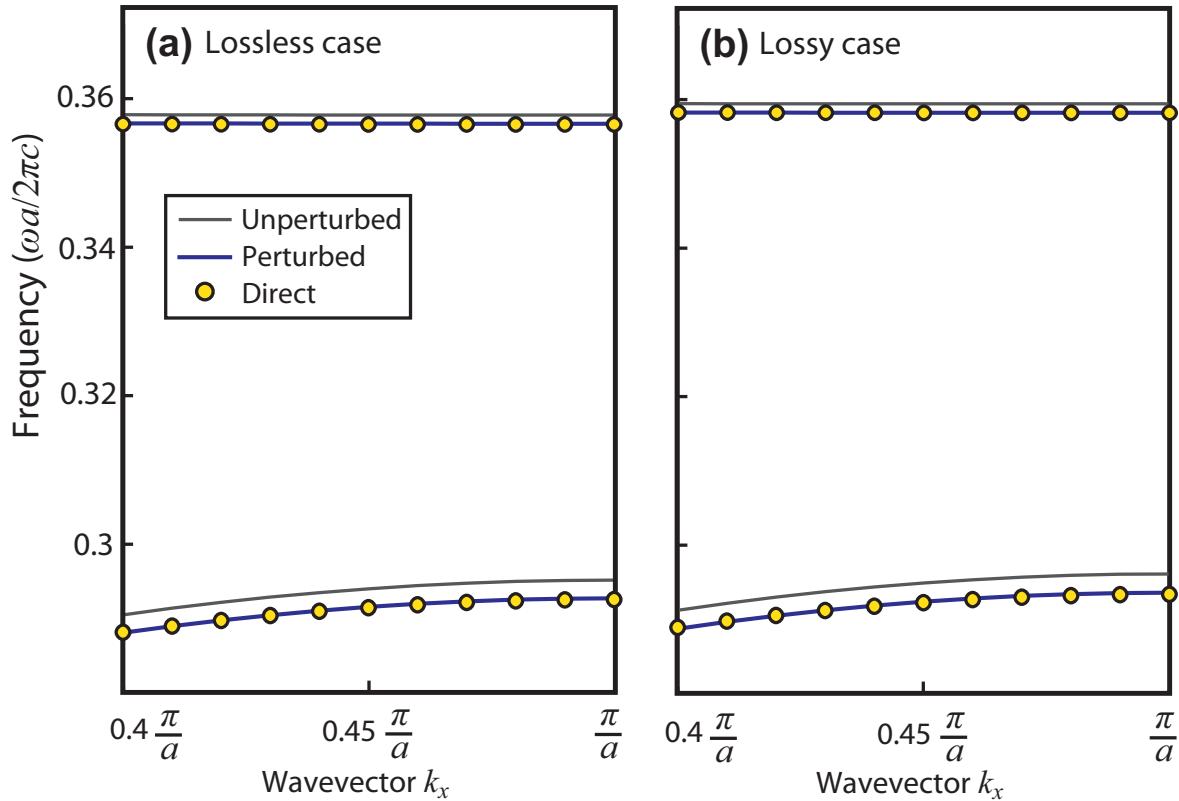}
  \end{center}
  \caption{Comparing the perturbation theory prediction ('Perturbed') and direct solution ('Direct') of $\omega_1$ in the region identified in Fig. 2(b), for dielectric $\Delta\varepsilon(\mathbf{r}) = 0.02$. The (a) lossless metal ($\Gamma_n = 0$) and (b) lossy metal cases shows excellent agreement for both the flat surface mode and non-surface mode. }
\end{figure}

\begin{figure}[t]
 \begin{center}
   \includegraphics[scale=0.63]{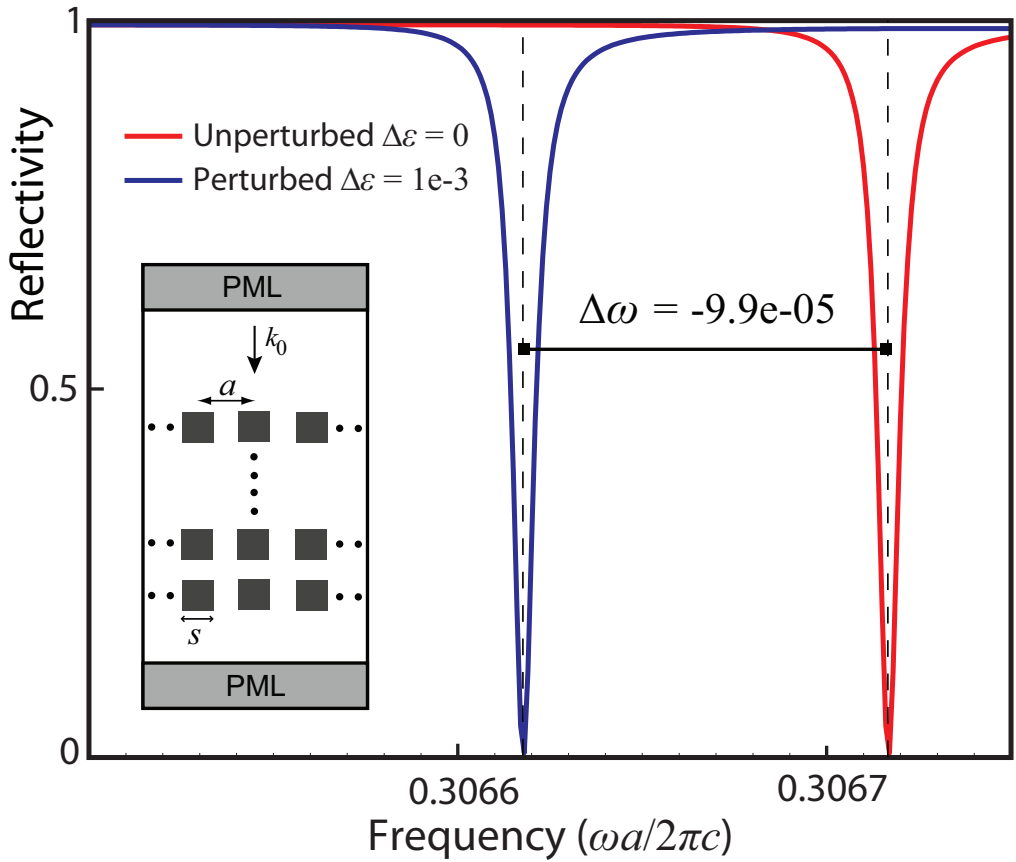}
  \end{center}
  \caption{Reflectivity spectrum of a finite 2D square lattice of square plasmonic rods ($s = 0.45a$) in air, consisting of 50 layers. The results are obtained with a full-field FDFD simulation (shown in the inset schematic) and show a reflectivity dip corresponding to the lowest-frequency propagating mode in the system (the mode at $\omega_c$ in Fig. 2(a)). Altering the air region by $\Delta\varepsilon = 1\mathrm{e}-3$ to simulate a perturbation, a shift in the dip of $\Delta\omega = -9.9\mathrm{e}-5$ is observed, matching the theoretical prediction of Eq. \eqref{refindlossless}, $\omega_1 = -9.37\mathrm{e}-5$, well. }
\end{figure}

To verify the accuracy of our perturbation theory approach we highlight a region of the TE band structure featuring both a surface and non-surface mode in Fig. 3. We alter the dielectric constant of the dielectric region by $\Delta\varepsilon = 0.02$, and calculate the resulting shift in eigenfrequency, using perturbation theory for both lossy and lossless cases. For the lossless system, we set $\Gamma_n = 0$, and use Eq. \eqref{refindlossless}. For the lossy system, we can use the exact perturbation theory result of Eq. \eqref{refindlossy}, as well as the approximation in terms of energy density in Eq. \eqref{lossyapprox}. The results from these two forms of perturbation theory are nearly identical to each other. These results, from both forms of the perturbation theory, are then compared to the band structure obtained by directly solving Eq. \eqref{maineqn} for the perturbed system. The results from the perturbation theory show excellent agreement with results from the direct calculation for both the lossless and lossy systems. 

\subsubsection{Practical Application}

As an illustration of the practical significance of this result, we analyze a potential refractive-index sensing scheme by calculating the reflection/transmission spectrum of 50-layers of the plasmonic rod system considered above, using a full-field 2D finite-difference frequency-domain (FDFD) simulation.\cite{veronis05} The plasmonic metal, corresponding as before to the two-pole fit of silver's dielectric function, is assumed to be lossless for the purposes of this illustration. For the TM case, below the cutoff frequency $\omega_c$ [Fig. 1(a)], no propagating modes are supported. Thus, the structure is strongly reflecting. The lowest-frequency dip in the reflection spectrum [Fig. 4] corresponds to the first mode supported by the system, identified previously in Fig. 2(a) at $\omega_c$. We can then examine the shift in this reflectivity dip when the dielectric surrounding the rods is changed by $\Delta \varepsilon$ (due to, for example, the introduction of a biochemical analyte). We observe a shift of the dip by $\Delta\omega = -9.9\mathrm{e}-5$ for $\Delta \varepsilon = 1\mathrm{e}-3$. Using Eq. \eqref{refindlossless} we calculate the shift of the lowest-frequency mode at $k = 0$, for the same $\Delta \varepsilon$, to be $\omega_1 = -9.37\mathrm{e}-5$, which matches well with the shift observed in the full-field simulation. Thus, the perturbation theory with respect to the eigenmodes of the system can be used to predict shifts in features of the same system's transmission and reflection spectra.

\section{Application II: Metallic Plasma Frequency Modulation}

In this section we demonstrate that we can also treat small changes in the plasma frequency of a metal present in a plasmonic nanostructure. This is of interest given a recent experimental work with ITO\cite{Diest:2010} where an applied electric field induced a change in carrier concentration thereby changing its plasma frequency and behavior at optical frequencies. In other recent work, the infrared plasmonic resonance of a metallic nanostructure (split-ring resonators) immersed in an electrolyte was affected by applying a voltage that altered the structure's average electron bulk density.\cite{Shao:2010} 

To derive an expression for the modal frequency change in such situations, without loss of generality, we will assume that only the $N$-th pole experiences a plasma frequency modulation. Then, we note that to first order the effect of $\Delta\omega_{p,N}$ affects our matrix elements by $\omega_{p,N}^{2} \rightarrow \omega^{2}_{p,N} + 2\omega_{p,N}\Delta \omega_{p,N}$. With this established we now separately consider the lossless and lossy cases.

\subsection{Lossless case, $\Gamma_n = 0$}

For the lossless case the following perturbing matrices, $\mathbf{A}_1$ and $\mathbf{B}_1$ are applicable:
\begin{equation}
\mathbf{A}_1 = \mathrm{diag} \left (0, 0,  \cdots, -\frac{\omega_{0,N}^2}{\omega_{p,N}^{2}\varepsilon_\infty}\left ( \frac{2 \Delta\omega_{p,N}}{\omega_{p,N}} \right ), -\frac{1}{\omega_{p,N}^{2}\varepsilon_\infty}\left ( \frac{2 \Delta\omega_{p,N}}{\omega_{p,N}} \right ) \right )
\end{equation} 
\begin{equation}
\mathbf{B}_1 = \left ( \begin{array}{cccc}
0 & \cdots & 0 & 0\\
\vdots & & \ddots & \vdots \\
0 & \cdots & 0 & i\frac{\omega_{0,N}^2}{\omega_{p,N}^{2}\varepsilon_\infty}\left ( \frac{2 \Delta\omega_{p,N}}{\omega_{p,N}} \right )\\
0 & \cdots & -i\frac{\omega_{0,N}^2}{\omega_{p,N}^{2}\varepsilon_\infty}\left ( \frac{2 \Delta\omega_{p,N}}{\omega_{p,N}} \right ) & 0 \end{array} \right ).
\end{equation}
Since the eigenvalue equation Eq. \eqref{maineqn} is Hermitian for the lossless case, Eq. \eqref{generalperturb} becomes:
\begin{equation}
\omega_1 = \frac{\mathbf{x}_{0}^{\dagger}\mathbf{B}_{1}\mathbf{x}_{0} - \omega_{0}\mathbf{x}_{0}^{\dagger}\mathbf{A}_{1}\mathbf{x}_{0}}{\mathbf{x}_{0}^{\dagger}\mathbf{A}_{0}\mathbf{x}_{0}}
\end{equation}
 which reduces to the following expression in terms of the constituent fields (as determined by solving Eq. \eqref{maineqn}):
\begin{equation}
\label{plasmalossless}
\omega_1 = \frac{\omega_0}{\int d\mathbf{r}~W_{0}}\int d\mathbf{r} 
\frac{ 2\Delta \omega_{p,N}(\mathbf{r})}{\omega_{p,N}} \left ( \frac{\omega_{0}^{2} - \omega_{0,N}^{2}}{\omega_{p,N}^{2}} \right ) |\mathbf{P}_{N}(\mathbf{r})|^{2}
\end{equation}

The frequency shift $\omega_1$ for a given mode is thus directly proportional to the strength of the modal mechanical field in the $N$-th pole that is undergoing modulation, and the proximity of the unperturbed modal frequency $\omega_0$ to $\omega_{0,N}$. Of particular interest is that, assuming $\Delta\omega_{p,N} > 0$, $\omega_1$ is positive or negative depending on whether the unperturbed modal frequency is above or below the Lorentz pole's resonance frequency $\omega_{0,N}$. We verify this behavior numerically in Section C below.

\subsection{Lossy case, $\Gamma_n \neq 0$}

As noted in the previous section, for the lossy case, Eq. \eqref{maineqn} is no longer Hermitian and $\mathbf{z}_0 \neq \mathbf{x}_0^{\dagger}$. Thus, Eq. \eqref{generalperturb} in this case reduces to:
\begin{align}
\label{plasmalossy}
\omega_1 = \frac{\mathbf{z}_{0}\mathbf{B}_{1}\mathbf{x}_{0}-\omega_{0}\mathbf{z}_{0}\mathbf{A}_{1}\mathbf{x}_{0}}{\mathbf{z}_{0}\mathbf{A}_{0}\mathbf{x}_{0}}
\end{align}
We note here that $\mathbf{B}_1$ is slightly altered in the lossy case:
\begin{equation}
\mathbf{B}_1 = \left ( \begin{array}{cccc}
0 & \cdots & 0 & 0\\
\vdots & & \ddots & \vdots \\
0 & \cdots & 0 & i\frac{\omega_{0,N}^2}{\omega_{p,N}^{2}\varepsilon_\infty}\left ( \frac{2 \Delta\omega_{p,N}}{\omega_{p,N}} \right )\\
0 & \cdots & -i\frac{\omega_{0,N}^2}{\omega_{p,N}^{2}\varepsilon_\infty}\left ( \frac{2 \Delta\omega_{p,N}}{\omega_{p,N}} \right ) & -i\frac{\Gamma_{N}}{\omega_{p,N}^{2}\varepsilon_\infty}\left ( \frac{2 \Delta\omega_{p,N}}{\omega_{p,N}} \right ) \end{array} \right ).
\end{equation}

As in Sec. IV.B, we note that empirically at optical frequencies $\mathbf{z}_0 \approx \mathbf{x}_0^{\dagger}$, and the denominator of Eq. \eqref{plasmalossy} $\sim\int d\mathbf{r}~W_{0}$. 

\begin{figure}[t]
 \begin{center}
   \includegraphics[scale=0.62]{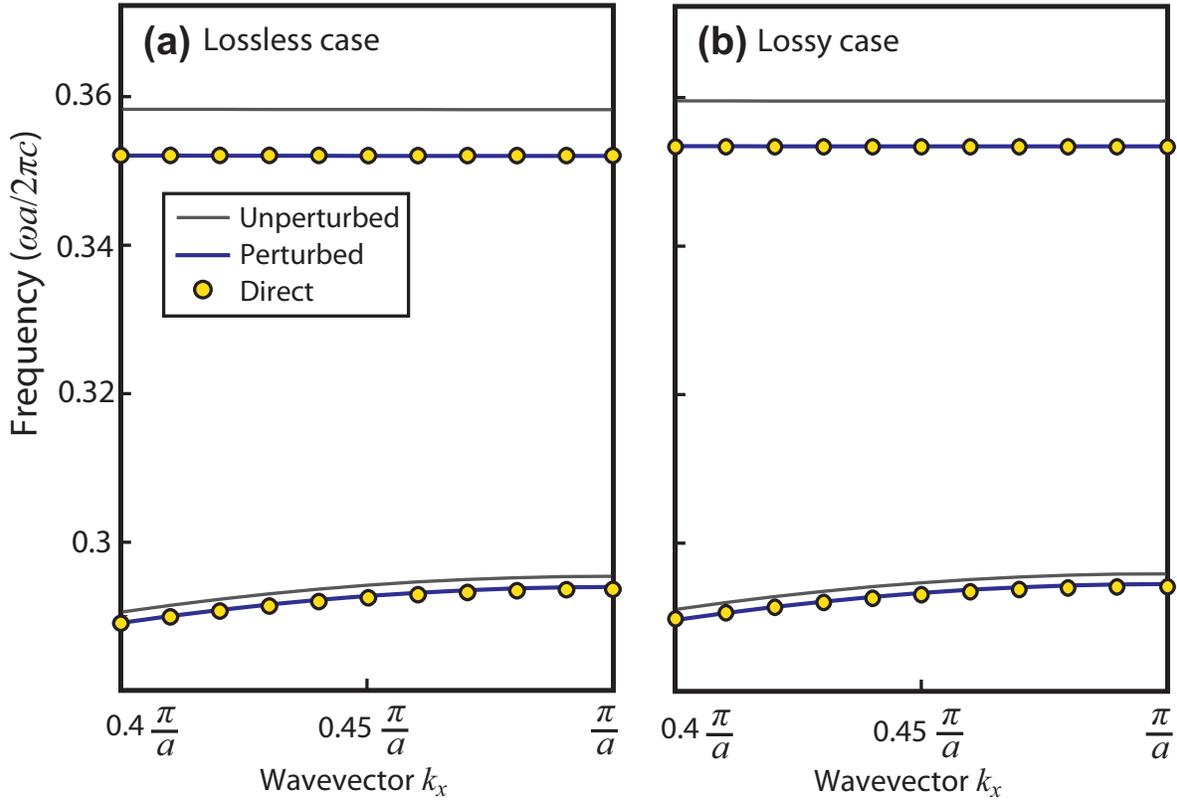}
  \end{center}
  \caption{Comparing the perturbation theory prediction ('Perturbed') and direct solution ('Direct') of $\omega_1$ in the same zoomed-in region identified in Fig. 2(b), for $\Delta\omega_{p,1} = 0.05 \omega_{p,1}$. The (a) lossless  ($\Gamma_n = 0$) and (b) lossy cases shows excellent agreement for both the flat surface mode and non-surface mode. Note the substantially greater $\omega_1$ for the flat plasmon mode which has a stronger mechanical component to its eigenmode, and that $\omega_1 < 0$ as predicted by Eq. \eqref{plasmalossless}.}
\end{figure}

\subsection{Numerical Example}

We consider the same example nanostructure of plasmonic rods in air used in the previous section, this time to numerically verify the accuracy of the perturbation theory for plasma frequency modulation in the plasmonic rod. Specifically, we focus again on the region of the TE band structure highlighted in Fig. 2(b) featuring both a surface and non-surface mode. As before, in addition to the original structure's TE bands, we directly calculate the bands of the perturbed structure, where the metal rod's Lorentz pole is modulated as $\Delta \omega_{p,1} = 0.05 \omega_{p,1}$. We now use the perturbation theory calculations to predict the effect of $\Delta \omega_{p,1}$ and compare it to the direct result. We note excellent agreement in Fig. 5 between the perturbation theory and direct approaches for both lossless and lossy cases. 

\begin{figure}[t]
 \begin{center}
   \includegraphics[scale=0.62]{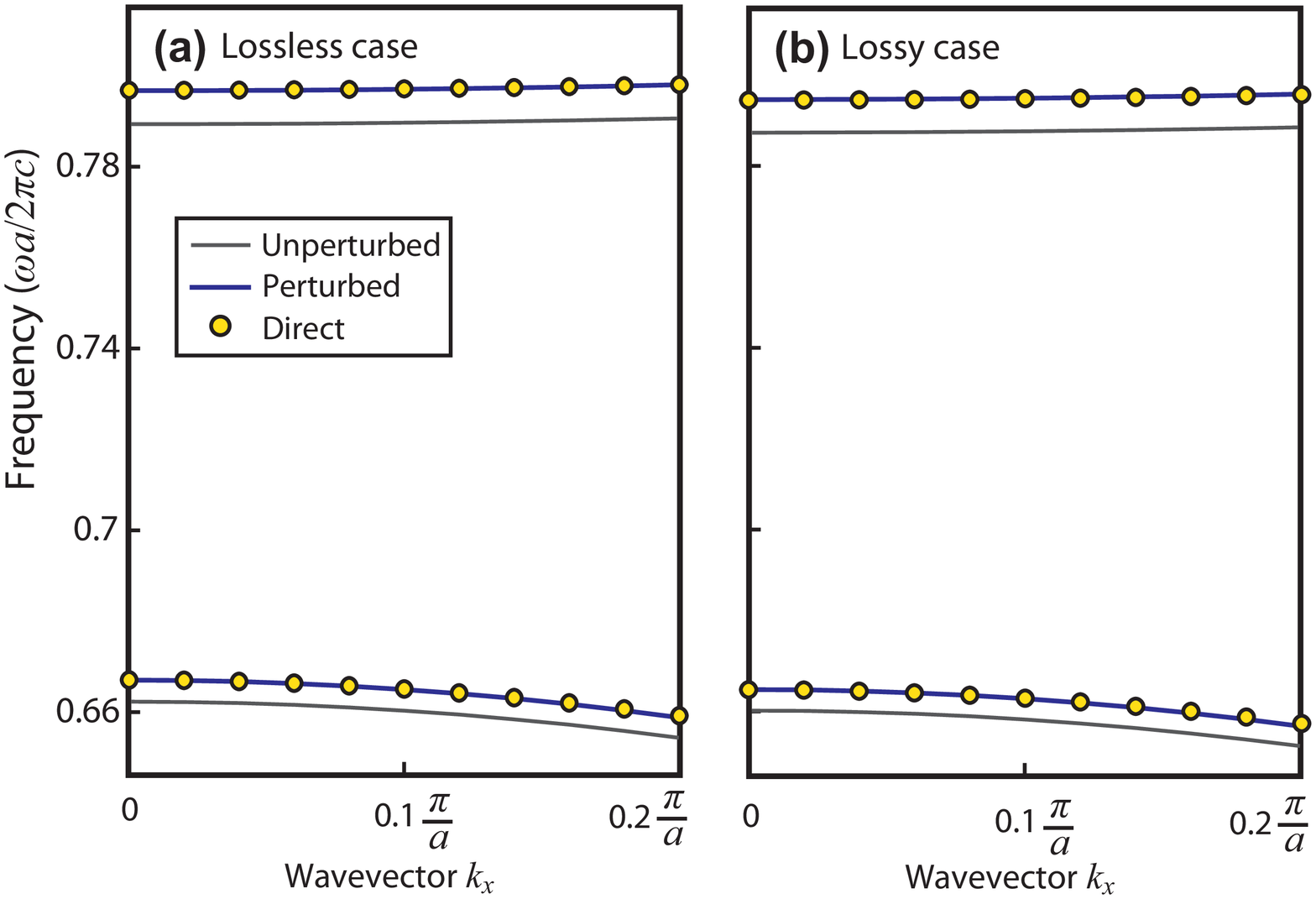}
  \end{center}
  \caption{Comparing the perturbation theory prediction ('Perturbed') and direct solution ('Direct') of $\omega_1$ in the region above $\omega_{0,1}$ identified in Fig. 2(b), for $\Delta\omega_{p,1} = 0.05 \omega_{p,1}$. The (a) lossless  ($\Gamma_n = 0$) and (b) lossy cases show $\omega_1 > 0$ (as opposed to $\omega_1 < 0$ in Fig. 5), verifying the prediction of Eq. \eqref{plasmalossless}.}
\end{figure}

Moreover, we note that in both lossy and lossless cases, the surface plasmon mode experiences a substantially greater $\omega_1$ compared to the lower frequency non-surface mode. This is consistent with the observation from Raman and Fan\cite{Raman:2010} that surface modes have stronger mechanical field intensities, and the form of Eqs. \eqref{plasmalossless} and \eqref{plasmalossy}, which predict greater $\omega_1$ for modes with stronger mechanical fields. This behavior is of practical interest for active plasmonic device applications.

Finally, as observed in Eq. \eqref{plasmalossless}, for a positive shift of the plasma frequency $\Delta \omega_{p,1} >0$, we expect either a positive or negative frequency shift $\omega_1$ depending on whether a mode's $\omega$ is greater or less than $\omega_{0,1}$. In Fig. 5 the modes presented are below $\omega_{0,1}$ and $\omega_1 < 0$, as expected for $\Delta\omega_{p,1} > 0$. In Fig. 6, we examine modes in the region above $\omega_{0,1}$ highlighted in Fig. 2(b), and note that $\omega_1 > 0$ as predicted by Eq. \eqref{plasmalossless}. This unusual property is highlighted here for the first time in the literature, and of potential utility as research in active plasmonics progresses.

\section{Conclusion}
We have developed and verified a perturbation theory that provides considerable insight into the effect of small variations in both the dielectric and metallic components of plasmonic and dispersive metamaterial nanostructures. Using a previously developed photonic band formalism for general dispersive nanostructures described by an arbitrary number of Lorentz poles, we are able to accurately predict the effect of modulation in the dielectric refractive index and metallic plasma frequency of a metal-dielectric plasmonic system. We specifically highlight new behaviors related to plasma frequency modulation that are of potential significance for future research in active plasmonics.

Given increasing interest in such dispersive systems for sensing and modulation applications, and active devices more generally, the theory developed above will indeed be of considerable utility in future designs and analyses.

\acknowledgements
This work was supported by the Center for Advanced Molecular Photovoltaics (CAMP) (Award No KUSC1-015-21), made by King Abdullah University of Science and Technology (KAUST), and the Interconnect Focus Center, funded under the Focus Center Research Program (FCRP), a Semiconductor Research Corporation entity.

\end{document}